\documentclass[journal=jacsat,manuscript=article]{achemso}

\usepackage[version=3]{mhchem} 
\usepackage[T1]{fontenc}
\usepackage{siunitx}

\author{Yanzhao Guo}
\altaffiliation{Translational Research Hub, Maindy Road, Cardiff, CF24 4HQ, United Kingdom}
\author{John P. Hadden}
\altaffiliation{Translational Research Hub, Maindy Road, Cardiff, CF24 4HQ, United Kingdom}
\author{Rachel N. Clark}
\altaffiliation{Translational Research Hub, Maindy Road, Cardiff, CF24 4HQ, United Kingdom}
\author{Samuel G. Bishop}
\altaffiliation{Translational Research Hub, Maindy Road, Cardiff, CF24 4HQ, United Kingdom}
\author{Anthony J. Bennett}
\affiliation{School of Engineering, Cardiff University, Queen’s Buildings, The Parade, Cardiff, CF24 3AA, United Kingdom}
\email{BennettA19@cardiff.ac.uk}
\phone{}
\fax{}
\altaffiliation{Translational Research Hub, Maindy Road, Cardiff, CF24 4HQ, United Kingdom}

\title[Photo-dynamics of quantum emitters in aluminum nitride]
  {Photo-dynamics of quantum emitters in aluminum nitride}

\abbreviations{}
\keywords{Aluminum nitride, quantum emitters, optical dynamics, charge}

\begin{document}


\begin{abstract}
Aluminum nitride is a technologically important wide bandgap semiconductor which has been shown to host bright quantum emitters. In this paper, we probe the photo-dynamics of quantum emitters in aluminum nitride using photon emission correlations and time-resolved spectroscopy. We identify that each emitter contains as many as 6 internal energy levels with distinct laser power-dependent behaviors.  Power-dependent shelving and de-shelving processes, such as optically induced ionization and recombination are considered, indicating complex optical dynamics associated with the spontaneous and optically pumped transitions. State population dynamics simulations qualitatively explain the temporal behaviours of the quantum emitters, revealing that those with pump-dependent de-shelving processes can saturate at significantly higher intensities, resulting in bright room-temperature quantum light emission.
\end{abstract}


\section{Introduction}

Single quantum emitters (QEs) in wide bandgap semiconductors are promising single-photon sources which can operate up to room temperature \cite{Aharonovich2017QuantumDimensions, Berhane2017BrightNitride, Atature2018MaterialTechnologies}. Compared with the well-known negatively-charged nitrogen-vacancy (NV) color center in diamond \cite{Doherty2013TheDiamond}, many of the QEs reported in III-nitride semiconductors show favourable optical properties \cite{Aharonovich2017QuantumDimensions} such as higher brightness  \cite{Guo2023CoherentTemperature.}, improved spectral purity \cite{Patel2022ProbingNitride} and potential industrial scalability \cite{Sajid2020Single-photonProgress}. Recently, QEs in hexagonal boron nitride (hBN) and gallium nitride (GaN) have been reported with optically detected magnetic resonance (ODMR) response \cite{Gottscholl2020InitializationTemperature, Stern2022Room-temperatureNitride, Luo2023RoomGaN}, which makes them attractive for quantum sensing and, potentially, spin-based quantum computation \cite{Gottscholl2021SpinSensors, Ramsay2023CoherenceNitride., Guo2023CoherentTemperature., Stern2023ATemperature}. Another member of the III-nitride semiconductor family,  aluminum nitride (AlN), also possesses various QEs which have been reported with low multi-photon emission rates ($g^{(2)}$(0)<0.1) \cite{Lu2020BrightPhotonics}, near 65\% Debye-Waller factor\cite{Wang2023QuantumLaser} and almost 1 MHz photon detection rates \cite {Bishop2022Evanescent-fieldLens, Xue2021ExperimentalFilms}. Moreover, theoretical calculations show that AlN QEs may host spin states with optically addressable transitions \cite{Varley2016DefectsQubits}.

However, potential applications require improved knowledge of the QE's internal electronic structure and rates of radiative and non-radiative transitions \cite{Patel2022ProbingNitride}. 
Previous studies on AlN QEs have reported the photon bunching associated with at least one metastable dark state ('shelving state') \cite{Bishop2022Evanescent-fieldLens, Xue2021ExperimentalFilms, Lu2020BrightPhotonics, Wang2023QuantumLaser}, and yet the transitions between these states remain unknown. Understanding the transitions between internal energy levels in single QE systems is an important step in the effort to unpick the physical origin of the QEs. It is also required to explain effects such as spin-pumping by a green laser, a first step in observing ODMR in quantum sensing experiments \cite{Stern2022Room-temperatureNitride, Luo2023RoomGaN}. 

In this paper, we use photon emission correlation spectroscopy (PECS), time-resolved photoluminescence (TRPL) and state population dynamics simulations to probe the photo-dynamics of emitters with differing behaviors. We infer that there are at least six internal energy levels, which govern the TRPL, bunching and saturation of the optical transition. We find two classes of QEs with different power-dependent shelving processes associated with charge ionization and recombination. These results demonstrate that photon bunching caused by shelving the system in a dark state inherently limits the saturation rate of the photon source. In emitters where increasing optical power de-shelves the dark state, we observe an increased photon emission intensity.

\section{Results and discussion}

We use a home-built confocal microscope to study isolated single QEs in a commercial single-crystal c-plane \SI{1}{\micro\metre} AlN film on a sapphire template at ambient conditions \cite{Bishop2020Room-TemperatureNitride}.  By investigating 10 QEs in this AlN film we identify two classes of QE in which shelving processes are found to increase or decrease with laser power, exemplified by emitters QE A and QE B, respectively. All the measurements are made with the laser polarization aligned to the QE's preferred absorption polarization angle \cite{Bishop2020Room-TemperatureNitride} with no polarizer in the collection path. The absorption and emission polarization characterization is available in the Supporting Information (SI). Further details on the system are given in the Experimental Methods. 

\begin{figure}
\includegraphics[width=\textwidth]{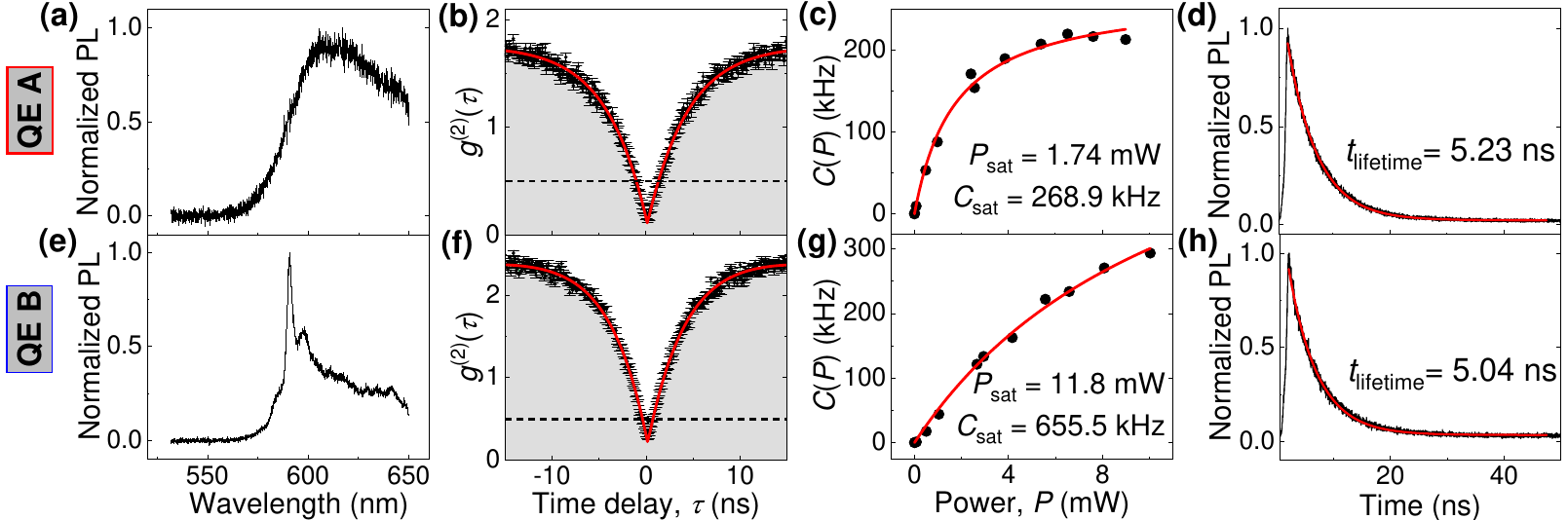}
\caption{Characterization of two quantum emitters in AlN at room temperature: QE A (first row) and QE B (second row). (The first column) (a) and (e) show the spectra between 532 and \SI{650}{\nano m}. (The second column) (b) and (f) are the photon emission correlation function presented normalised without background correction (black points) fit using an empirical model discussed in the text (red line). Error bars represent Poissonian uncertainties based on the photon counts in each bin. (The third column) (c) and (g) are the CW-PL saturation behaviors (black points) as a function of laser power, fit using an empirical saturation model equation \ref{eq:Psat}. (The last column) (d) and (h) show the excited state lifetime measurement under a \SI{520}{\nano m} pulsed laser excitation, fitted with a single exponential (red line).}
\label{Fig. 1}
\centering
\end{figure}

In Fig. \ref{Fig. 1}(a), the spectrum of QE A has a broad phonon sideband (PSB) from \SI{600}{\nano m} extending beyond the optical filtering cut-off at \SI{650}{\nano m}, suggesting a low Debye-Waller Factor. This is consistent with some previous reports \cite{Bishop2020Room-TemperatureNitride, Bishop2022Evanescent-fieldLens}. In contrast, in Fig. \ref{Fig. 1}(e) QE B displays a strong zero phonon line (ZPL) at \SI{590}{\nano m} with a PSB more comparable to other recent studies \cite{Wang2023QuantumLaser}. We speculate these two QEs originate from the same crystal complex but with different local strain and charge environments. Correlation histograms display substantial bunching over hundreds of nanoseconds, but nevertheless the values of $g^{(2)}$(0) for QE A and QE B are 0.16 (0.042) and 0.29 (0.034) respectively in Figs. \ref{Fig. 1}(b) and (f), confirming they are single QEs. Another feature of a quantized emitter is its photoluminescence (PL) intensity saturation with continuous wave (CW) laser power, shown in Figs. \ref{Fig. 1}(c) and (g), and fitted with, 
\begin{equation}
C(P)= \frac{{C_{\mathrm{sat}}}P}{{P+P_{\mathrm{sat}}}} \
\label{eq:Psat}
\end{equation}
where \emph{C(P)} is the steady-state PL rate as a function of power, $C_{\mathrm{sat}}$ is the saturation PL rate and $P_{\mathrm{sat}}$ is the corresponding saturation power. QE B requires 6.8 times higher $P_{\mathrm{sat}}$ and has 2.4 times higher $C_{\mathrm{sat}}$. Despite the difference in saturation behavior, the two QEs both have a $\sim$ \SI{5}{\nano s} radiative lifetime obtained by fitting a single exponential decay function in Figs. \ref{Fig. 1}(d) and (h), suggesting the difference in saturated behavior is a result of differing non-radiative pathways \cite{Kianinia2018All-opticalMaterials., Han2012DarkDiamond}. 
 
\begin{figure}
\includegraphics[width=\textwidth]{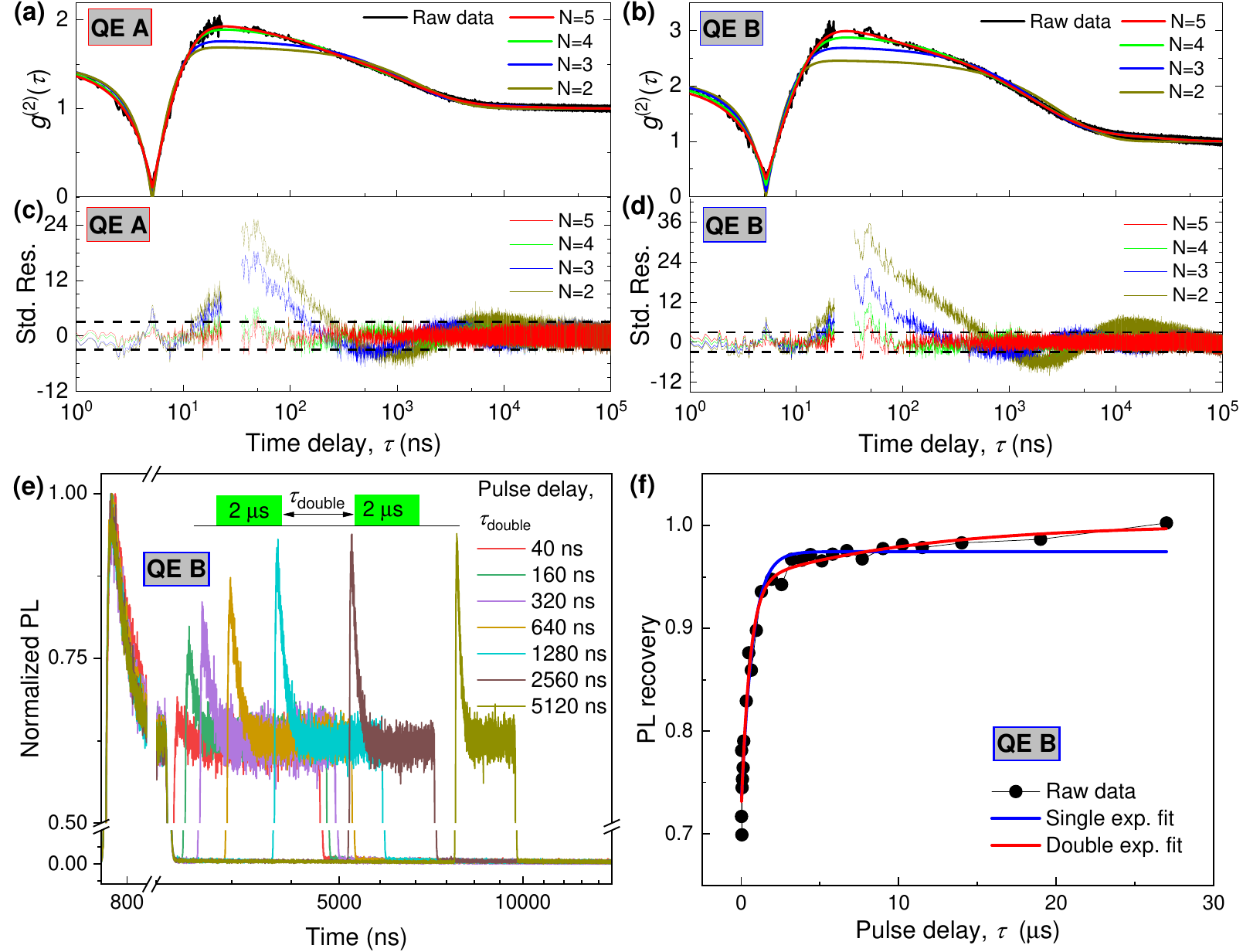}
\caption{Photon emission correlation spectroscopy and double pulse laser excitation. Black lines in (a) and (b) are the PECS of QE A and QE B, as well as their fitting curve by modes 5, 4, 3, and 2 of the empirical Equation \ref{eq:g2_empirical}. (c) and (d) are the corresponding residuals from the fits to QE A and QE B, respectively. The raw data between 22-\SI{35}{\nano s} is masked to hide reflections from the detector's backflash.  (e) is the TRPL of QE B under double pulse laser excitation. The inset is the train of the laser pulses. (f) represents the PL revival behavior under the second pulse excitation in (e) fitted by single exponential and double exponential equations.}
\label{Fig. 2}
\centering
\end{figure}

To further explore the dynamics of photon emission in these QEs, PECS was recorded for QE A and QE B (Figs. \ref{Fig. 2}(a) and (b)) over the \SI{100}{\pico s} to \SI{100}{\micro s} time scale. Least-squares fits to the $g^{(2)}$(\(\tau\)) data is shown using the empirical equation, 
\begin{equation}
g^{(2)}(\tau)= 1-C_1e^{-|\tau-\tau_0|\cdot r_1}+\sum^N_{i=2
}C_ie^{-|\tau-\tau_0|\cdot r_i} \
\label{eq:g2_empirical}
\end{equation}
with varying numbers of variables in the sum, denoted $i$. As we shall show later, the total number of levels in the sum indicates the number of shelving states in the QE, \emph{N}-1. Here, $\tau_0$ is the delay time offset of the two detectors, $r_{1}$ is the antibunching rate, $C_{1}$ is the antibunching amplitude, $r_{i}$ for \emph{i} \(\geq\) 2 are bunching rates, and $C_{i}$ for \emph{i} \(\geq\) 2 are the corresponding bunching amplitudes. Then the number of resolvable timescales, \emph{N}, has been determined by calculating and comparing the reduced chi-squared statistic, r-square for each best-fit model. Figs. \ref{Fig. 2}(c) and (d) show standardized residuals for each QE for the best-fit empirical model at different \emph{N}. Interestingly, we observe that \emph{N} = 5 is best able to match our results, due to the obvious deviation at \(\tau\) = $10^1$-$10^3$ ns for \emph{N} = 4, 3, 2. The number (\emph{N}) of observed timescales ranging from \SI{}{\nano s} to tens of \SI{}{\micro s} represents at least \emph{N}-1 shelving states, which is large compared with the previous reports \cite{Bishop2020Room-TemperatureNitride}. Some states could represent multiplets associated with different spin manifolds \cite{Patel2022ProbingNitride, Stern2022Room-temperatureNitride} or this could be a result of fluorescence intermittency caused by charging of nearby trap sites \cite{Davanco2014MultipleSources, Efros2016OriginDots, Frantsuzov2008UniversalNanowires}.

To further verify the shelving state dynamics, we have excited QE B by double pulses \SI{2}{\micro s} in length at \SI{532}{\nano m} with variable spacing, $\tau_{\mathrm{double}}$ (see inset of Fig. \ref{Fig. 2}(e)). The first pulse results in a quasi-steady state by pumping the population into the shelving states. Dependent on the delay between the two pulses we observe a revival of the PL emission under the second pulse excitation, as the population decays back to the ground state from the shelving states in Fig. \ref{Fig. 2}(e) \cite{Robledo2011SpinDiamond}. Integrating the first \SI{120}{\nano s} PL at the start of the second pulse, we plot the PL revival curve in Fig. \ref{Fig. 2}(f). The double exponential gives an adequate fit, indicating more than one decay rate associated with the shelving states. This result further supports the $g^{(2)}$(\(\tau\)) fitting’s observation indicating multiple shelving dynamic processes. The fact this data can be modelled with 2 shelving states when there is no optical power during $\tau_{\mathrm{double}}$, rather than the 4 shelving states required for $g^{(2)}$(\(\tau\)), may be a result of some laser-driven de-shelving mechanism in QE B.

\begin{figure}
\includegraphics[width=\textwidth]{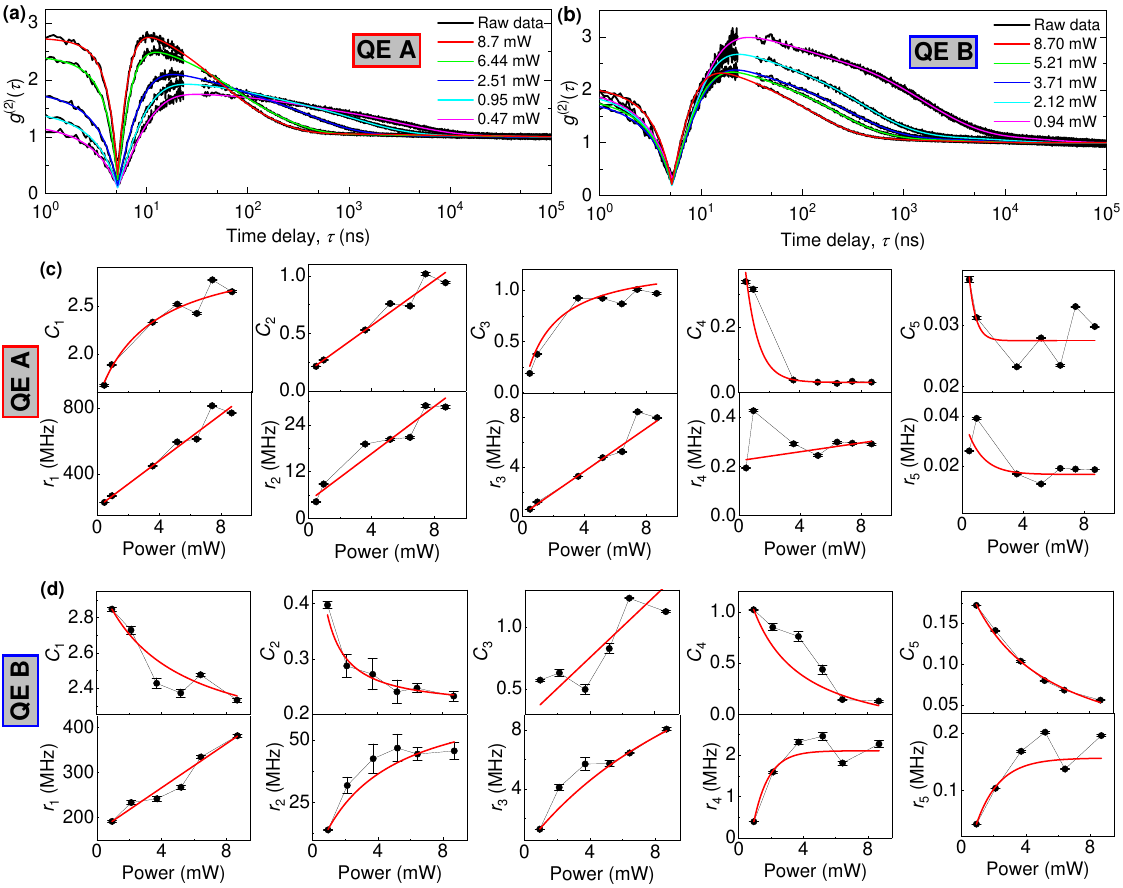}
\caption{Power-dependent PECS from (a) QE A and (b) QE B. Each autocorrelation is fitted with Equation \ref{eq:g2_empirical} and \emph{N} = 5. (c) and (d) are amplitudes $C_i$ and rates $r_i$ arising from fitting the Equation \ref{eq:g2_empirical} for QE A and QE B, respectively.}
\label{Fig. 3}
\centering
\end{figure}

To investigate the power-dependent dynamics, the power-dependent $g^{(2)}$(\(\tau\)) data has been fitted with Equation \ref{eq:g2_empirical} for \emph{N} = 5. In Figs. \ref{Fig. 3}(a) and (b), QE A and QE B show nearly opposite power-dependent bunching mechanisms, indicating the different power-dependent shelving dynamics. Figs. \ref{Fig. 3}(c) and (d) summarize the fitting results of the antibunching and bunching rates and amplitudes for QE A and QE B, respectively, with red lines as a guide to the eye. 
 
For QE A, the dominant $C_{2,3}$ bunching amplitudes rise with laser power, thus the bunching increases in Fig. \ref{Fig. 3}(a). This trend reveals that the increasing laser power transfers the population from the excited state to the shelving states, reducing the PL intensity \cite{Han2012DarkDiamond}. This power-enhanced bunching behaviour is consistent with previous reports \cite{Bishop2020Room-TemperatureNitride, Bishop2022Evanescent-fieldLens}. In contrast, for QE B, the bunching amplitudes $C_{2,4,5}$ fall with power, resulting in a net reduction in bunching with increasing laser intensity. In other words, increasing laser power transfers the population out of the shelving states. This enables QE B to be an efficient radiative QE at high laser power \cite{Kianinia2018All-opticalMaterials., Khatri2020OpticalNitride, Han2012DarkDiamond}. Such behaviour has not been observed in AlN yet, but reveals how some QEs emitting in the same spectral range can exhibit differing photon bunching behaviour as a result of different internal energy levels and dynamics.

\begin{figure}[t]
\includegraphics[scale=0.8]{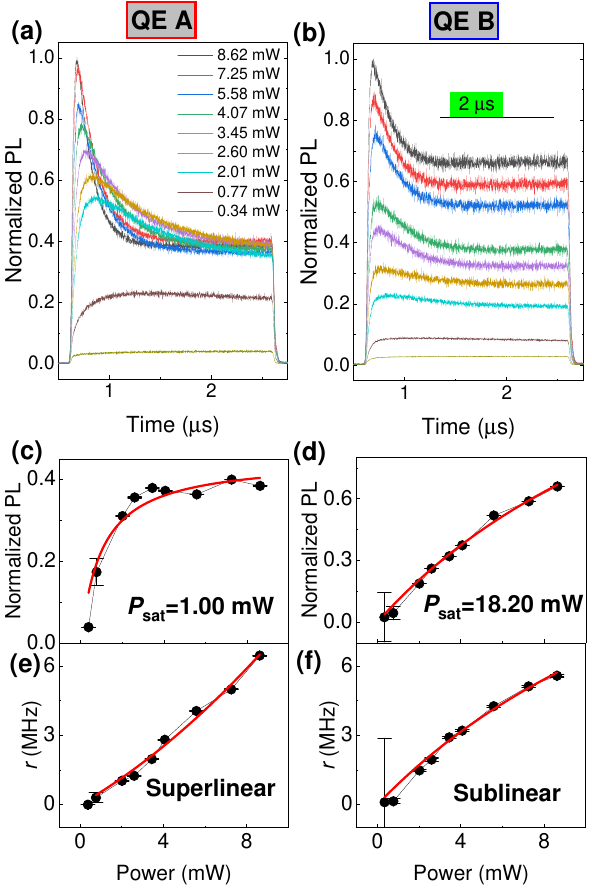}
\caption{Power-dependent TRPL with square excitation pulses. (a) QE A, (b) QE B, with insert showing the pulse sequence. (c) and (d) are the normalized PL saturation behaviours of the steady states. (e) and (f) are the TRPL decay rates observed in (a) and (b).}
\label{Fig. 4}
\centering
\end{figure}

The antibunching rates for QE A and QE B  scale linearly with power, indicating a single excited state\cite{Patel2022ProbingNitride, Xue2021ExperimentalFilms}. For QE A bunching rates $r_{2,3,4}$ show linear scaling with laser power, which arises when the laser drives a transition between the radiative states and the shelving state (e.g. via charge ionization or re-conversion) \cite{Patel2022ProbingNitride, Han2012DarkDiamond}. Regarding QE B, the bunching rates are more complicated with non-linear and zero offset behaviour, possibly as the transitions can occur both spontaneously and through the optically driven transitions between shelving states\cite{Patel2022ProbingNitride, Han2012DarkDiamond, Hacquebard2018Charge-stateDiamond}, as we will show later. 

To further verify the power-dependent optical dynamics, we record the power-dependent TRPL of QE A and QE B under \SI{2}{\micro s} square pulsed excitation in Figs. \ref{Fig. 4}(a) and (b). We fit the TRPL with a single exponential decay function to extract a decay rate and normalized steady PL rate in Figs. \ref{Fig. 4}(c)-(f). In Figs. \ref{Fig. 4}(c)  and (d), QE B has an 18.2 times higher saturation power than QE A. The discrepancy between this value and the ratio obtained from CW saturation (Figs. \ref{Fig. 1}(c) and (g))  may be a result of using short \SI{2}{\micro s} pulses, which does not allow enough time for the longer time-scale decay processes to reach equilibrium. These saturation behaviours also reveal that the population of the excited state in QE A is rapidly shelved at high power leading to reduced radiative emission. In contrast, QE B remains an effective emitter at high power. The shelving rate of QE A is super-linear and that of QE B is sub-linear. Referring to the example of NVs in diamond and QEs in hBN, these decay rates and saturation behaviors may originate from optically pumped shelving (e.g., charge ionization and conversion)\cite{Han2012DarkDiamond, Hacquebard2018Charge-stateDiamond, Patel2022ProbingNitride}.

\begin{figure}
\includegraphics[width=\textwidth]{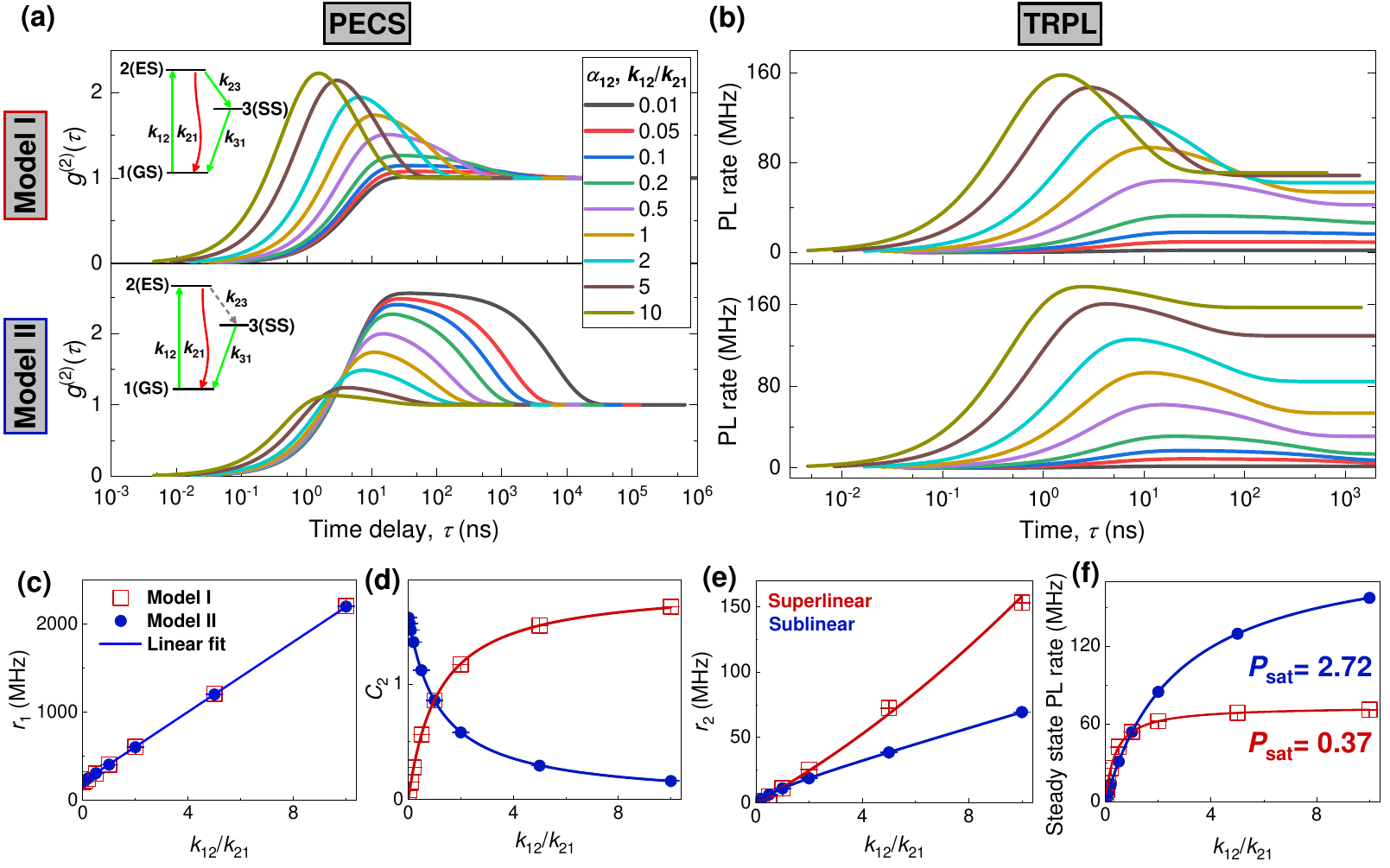}
\caption{State population dynamics simulation. (a) PECS and (b) TRPL simulation results for shelving Model I and Model II with transition rates from Table~\ref{tbl:notes1}. The inset in (a) is the proposed three-energy level shelving Model I and II which includes radiative emission (red arrow), optically pumped transitions (green arrows), and  non-radiative transitions with a fixed spontaneous decay rate (dotted grey arrows).  (c), (d) and (e) are the best-fit parameters $r_1$, $C_2$, and $r_2$ determined by fitting simulated $g^{(2)}$(\(\tau\)) data using Equation \ref{eq:g2_empirical} with \emph{N} = 2. (f) is the steady state PL saturation behaviour fitted by Equation \ref{eq:Psat}. The results are plotted as a function of  $k_{12}$/$k_{21}$, where  $k_{21}$= \SI{200}{\mega Hz} is a fixed parameter.}
\label{Fig. 5}
\centering
\end{figure}

To qualitatively understand these power-dependent behaviours, we calculate the PECS and TRPL for two different shelving models using a state population dynamics simulation in Figs \ref{Fig. 5}(a) and (b)\cite{Fishman2023Photon-Emission-CorrelationDefects}. For simplicity, we perform this simulation with three energy levels and different power-dependent rates in TRPL and PECS. We note that there are other possible models containing a single shelving level with power-dependent rates (See SI) and that these models could be extended to include all 4 shelving states inferred from Fig. \ref{Fig. 2}. However we show that the two single shelving models we consider are sufficient to reveal the physics of the power-dependent transitions, and to provide qualitative agreement with our experimental results.   Information regarding these simulations is given in the Experimental Methods. Each model consists of a ground state (GS) 1, an excited state (ES) 2, and a shelving state (SS) 3, where transitions between the states are labelled $k_{ij}$, where $i$ and $j$ are the initial and final state numbers in insert of  Fig. \ref{Fig. 5} (a). The transition rates of the two models are shown in Table~\ref{tbl:notes1}, but briefly, in Model I we assume both shelving and de-shelving transitions are driven by the laser, whereas in Model II there is a fixed shelving rate and optically pumped de-shelving rate. Figs. \ref{Fig. 5}(c), (d), and (e) show the results fitting these two models to PECS measurements with Eq. \ref{eq:g2_empirical} for \emph{N} = 2. The steady-state PL saturation is also simulated in Fig. \ref{Fig. 5}(f) by fitting the TRPL simulation with a single exponential equation. Moreover, we note that Fig. \ref{Fig. 5}(e) also represents the shelving rate derived from the TRPL simulation, attributed to the presence of a single shelving state in these three-energy level models.

 \begin{table}
  \caption{Simulation rate table}
    \label{tbl:notes1}
  \begin{tabular}{llll}
    \hline
    Model&Shelving rate & De-shelving & Emission transition rate \\
    \hline
    I &  $k_{23}$= 0.05*$k_{12}$\textsuperscript{\emph{a}}&$k_{31}$= 0.03*$k_{12}$\textsuperscript{\emph{a}}&$k_{21}$= 200 MHz  \\
    II  & $k_{23}$= 10 MHz&$k_{31}$= 0.03*$k_{12}$\textsuperscript{\emph{a}}&$k_{21}$= 200 MHz \\
 \hline
 \end{tabular}
 
\textsuperscript{\emph{a}} Optically excited rate $k_{12}$ = $\alpha_{12}$*$k_{21}$, $\alpha_{12}$=0.01, 0.05, 0.1, 0.2, 0.5, 1, 2, 5, 10.
\end{table}

In Fig. \ref{Fig. 5}(c) the antibunching rates ($r_{1}$) of these two models are linearly rising with pump power, offset by the spontaneous emission rate,  which is consistent with the results of QE A and QE B. In contrast, Model I and Model II display opposite power-dependent bunching dynamics. Specifically, Model I shows an increasing bunching amplitude comparable to what is observed in QE A.  On another hand, Model II shows a reduced bunching amplitude comparable to what is seen in QE B.  Moreover, interestingly, the TRPL shelving rates of Model I and Model II in Fig. \ref{Fig. 5}(e) can be fitted by superlinear and sublinear functions with zero offsets, which is perfectly consistent with the TRPL results of QE A and QE B in Figs. \ref{Fig. 4}(e) and (f). Additionally, Model II shows several times higher saturation power and saturation intensity than Model I in Fig. \ref{Fig. 5}(f), which is comparable to the results of QE A and QE B in Fig. \ref{Fig. 1}(c) and (g).

Thus, comparing our simulation and experimental results, we conclude that QE A displays Model I shelving and QE B displays Model II behaviour. We note that $r_{1}$ and $r_{4}$ for QE A have a non-zero value at low power due to spontaneous emission, and yet increase linearly with power suggesting some optical pumping is possible. In contrast, in QE B $r_{3}$, $r_{4}$ and $r_{5}$  are zero at low power (no decay by spontaneous emission from state 2) but saturate at high power, suggesting they can be optically pumped.

Based on the discussions above, a key factor for the achievable PL rate of QEs is the shelving dynamics at high power. The ideal QE should have reduced shelving at high power, as observed in QE B. We hypothesise that a second color laser could be used to efficiently repump the population from the shelving state back to the bright transition. In the best case, the non-radiative transition could be completely neglected, leading to an emitter with an intensity determined only by the spontaneous decay rate. For example, QE A would become $\sim$2.5 times brighter at saturation, giving > 0.67 MHz PL rate, and QE B would become 1.5 times brighter leading to $\sim$ 1.0 MHz PL rate.

\section{Conclusion}

In conclusion, AlN QEs display complex optical dynamics which indicates they have internal electronic level structures with multiple charge or shelving states. We identify two different optical-power-dependent shelving behaviours associated with the charge ionization and recombination processes. We propose models of the dynamic behaviour which complements previous reports and explains the qualitative features of our observations. Future experiments could focus on the energy-dependent behaviour of the shelving and de-shelving processes using tunable lasers. Nevertheless, the techniques used in this paper offer a way to study the internal energy levels in the QEs of other materials. Moreover, this study will help us to design a suitable protocol to minimise the time each QE spends in metastable shelving states, resulting in an overall increased intensity.

\section{Experimental Methods}

\subsection{Experiment }

The sample was excited by a CW \SI{532}{\nano m} laser (Crystal Laser) modulated by an acoustic-optic modulator (AOM) (ISOMET 553F-2) with < \SI{10}{\nano s} rise and fall time for static PL characterization, PECS and TRPL experiment. A \SI{100}{\pico s} pulsed \SI{520}{\nano m} laser (Picoquant P-C-520M) was used for the radiative lifetime measurement in Figs. \ref{Fig. 1}(d) and (h). The polarization of both lasers was set by a linear polarizer and half-waveplate. Excitation and collection of photons from the sample were performed by a single objective with NA=0.9. Collected PL was filtered by a dichroic mirror, \SI{532}{\nano m} long-pass filter and \SI{650}{\nano m} short-pass filter, before detection on SPCM-AQRH silicon avalanche photodiodes (Excelitas) or a spectrometer with a silicon CCD.

TRPL was recorded with an ID900 time controller. For the lifetime measurements (Figs. \ref{Fig. 1}(d) and (h)), the ID900 time controller records the PL histogram with a resolution of \SI{13}{\pico s} at \SI{20}{\mega Hz} repetition frequency. For the double-pulse laser excitation and single-pulse TRPL (Fig. \ref{Fig. 4}), the histogram was binned with \SI{1}{\nano s} resolution. The spacing between each laser pulse train is \SI{50}{\micro s} to reset the ground state population of QEs.

PECS was recorded using the ID900 time-tagging mode, with photon arrival times acquired from two detectors in a Hanbury-Brown and Twiss interferometer. Custom software numerically correlates each photon detection on one detector with all the other registered photons on the second detector, within a specified time window. 

PECS and TRPL data are presented normalised and without background correction. The spectra and saturation data in Fig. \ref{Fig. 1} are corrected by subtraction of background emission estimated by measurements from a location \SI{1}{\micro m} from the QE (See SI for the raw data).  In terms of the total system efficiency for an in-plane dipole, we consider the optical collection efficiency of the objective with NA=0.9 (4\(\%\))\cite{Bishop2022Evanescent-fieldLens},  the fiber coupling efficiency (38\(\%\)), and the detection efficiency of the single-photon detector (70\(\%\)). Therefore, we estimate the total system  efficiency is $\sim$ 1\(\%\). Additionally, we use a single NV center in bulk diamond as a reference to benchmark our experiment's excitation and collection performance (See SI). 

\subsection{Simulation}
For any \textit{N}-level electronic structure, the full optical dynamics are calculated by a system of \textit{N}-coupled differential equations \cite{Fishman2023Photon-Emission-CorrelationDefects}. 
\begin{equation}
\frac{dP}{dt}=G\cdot P
\end{equation}
where \textit{P} is a vector of state occupation probabilities and \textit{G} is the transition rate matrix, where the $G_{ij}$ represents the transition rate from \textit{i} state to \textit{j} state ($i\neq j$).  Each diagonal element $G_{ii}$ corresponds to the sum of all transition rates out of state $i$.

Then the autocorrelation function is proportional to the population of the radiative state, $P_2$($t_2$), given the system started in ground state $P_1$ following the detection of a photon at $t_1$, and then normalizing by the steady-state population of $P_2(\infty)$ \cite{Berthel2015PhotophysicsNanocrystals}.  This is given by
\begin{equation}
g^{(2)}(\tau)=\frac{P_2(t_2|P(t_1))}{P_2(\infty)}
\end{equation}
where \(\tau\) = $t_2$ - $t_1$ is the time delay of $g^{(2)}(\tau)$.

\begin{acknowledgement}
The authors acknowledge financial support provided by EPSRC via Grant No. EP/T017813/1 and EP/03982X/1 and the European Union's H2020 Marie Curie ITN project LasIonDef (GA No. 956387). RC was supported by grant EP/S024441/1, Cardiff University and the National Physical Laboratory. Sample processing was carried out in the cleanroom of the ERDF funded Institute for Compound Semiconductors (ICS) at Cardiff University.
\end{acknowledgement}

\section{Supporting Information}.


\setcounter{figure}{0}
\renewcommand{\thefigure}{S\arabic{figure}}
\begin{figure}
\includegraphics[scale=0.4]{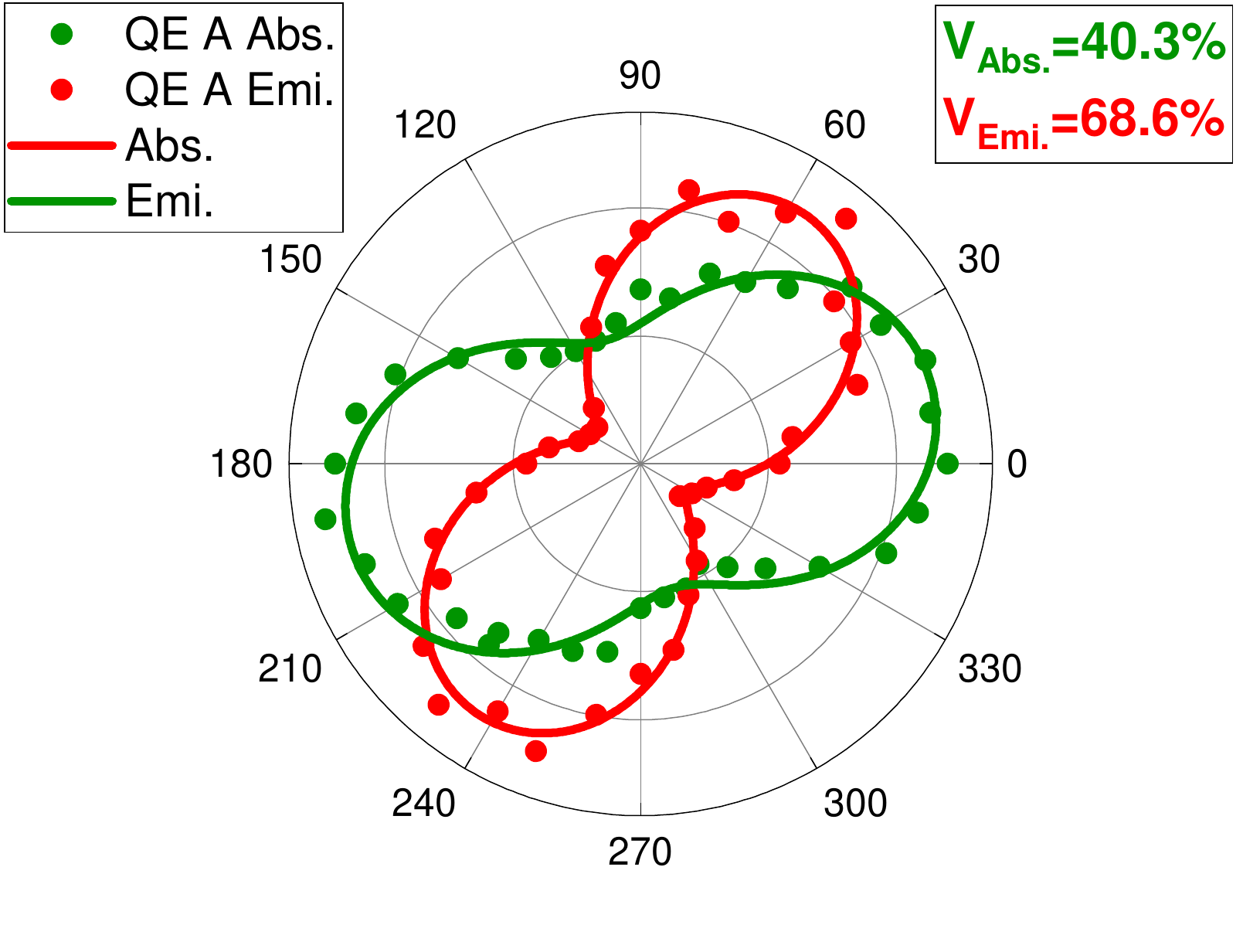}
\caption{Absorption and emission polarization plot of QE A.}
\label{Fig. S1}
\centering
\end{figure}
\setcounter{figure}{1}
\renewcommand{\thefigure}{S\arabic{figure}}
\begin{figure}
\includegraphics[width=\textwidth]{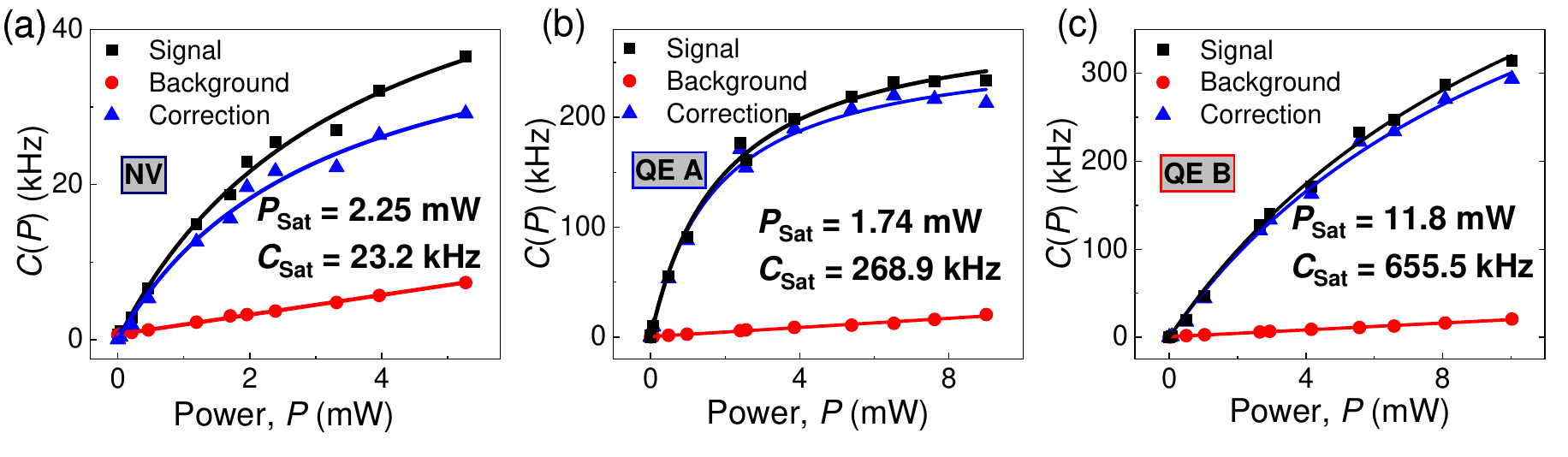}
\caption{A single NV center in diamond PL characterization compared with QE A and QE B.}
\label{Fig. S2}
\centering
\end{figure}

 \setcounter{table}{0}
\renewcommand{\thetable}{S\arabic{table}}
\begin{table}
  \caption{Simulation rate table}
    \label{tbSl:notesS1}
  \begin{tabular}{llll}
    \hline
    Model&  Shelving & De-shelving & Emission transition \\
    \hline
    III&  $k_{23}$= 10 MHz&$k_{31}$= 4 MHz&$k_{21}$= 200 MHz  \\
    IV& $k_{23}$= 10 MHz&$k_{31}$= 4 MHz, $k_{32}$=0.005*$k_{12}$\textsuperscript{\emph{a}}&$k_{21}$= 200 MHz \\
 \hline
 \end{tabular}
 
\textsuperscript{\emph{a}} Optically excited rate $k_{12}$ = $\alpha_{12}$*$k_{21}$, $\alpha_{12}$=0.01, 0.05, 0.1, 0.2, 0.5, 1, 2, 5, 10.
\end{table}

\setcounter{figure}{2}
\renewcommand{\thefigure}{S\arabic{figure}}
\begin{figure}
\includegraphics[width=\textwidth]{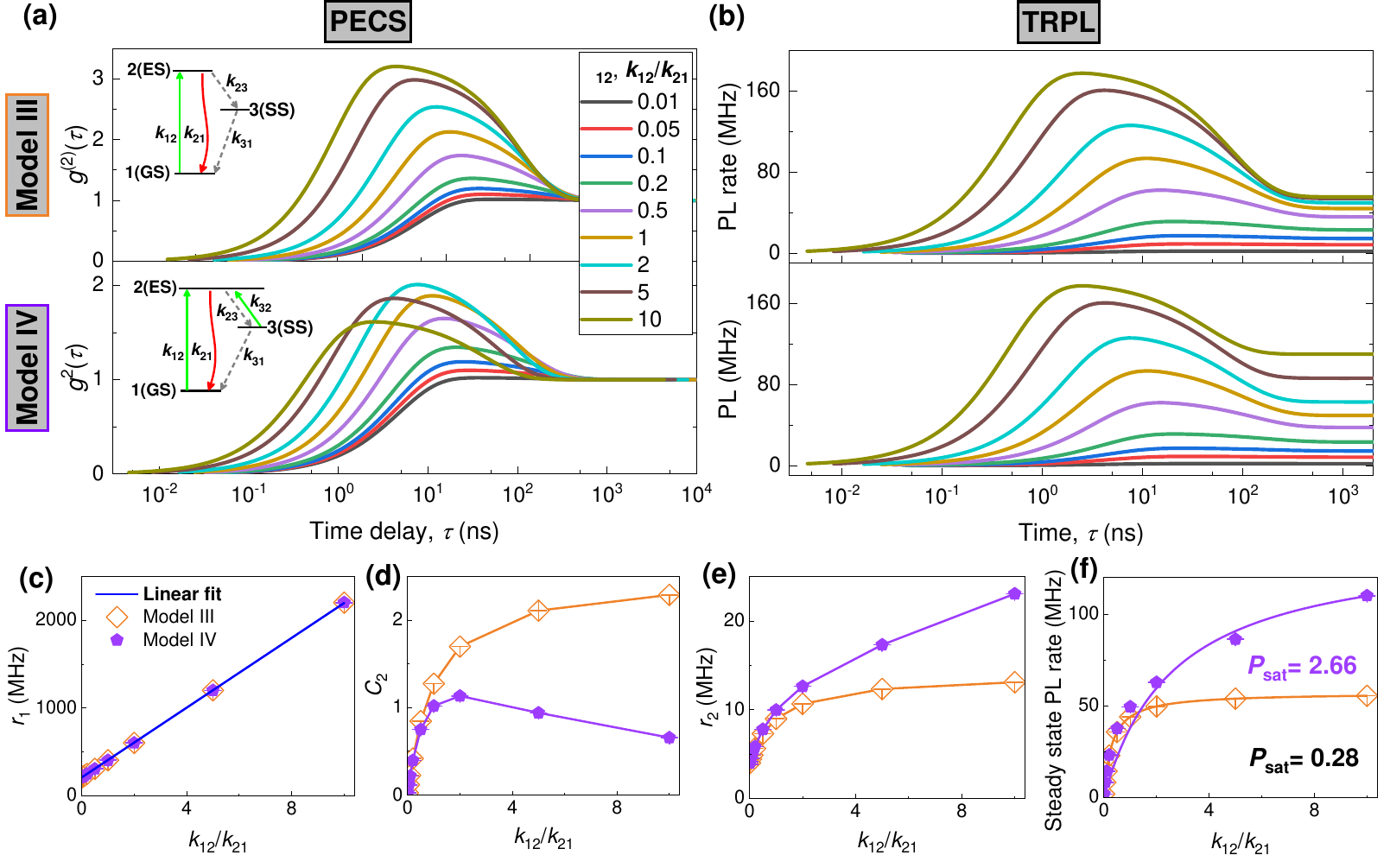}
\caption{State population dynamics simulation of extra two Models. (a) PECS and (b) TRPL simulation results for shelving model III and model IV with transition rates from Table~\ref{tbSl:notesS1}. The inset in (a) is the proposed three-energy level shelving model III and IV which includes radiative emission (red arrow), optically pumped transitions (green arrows), and  non-radiative transitions with a fixed spontaneous decay rate (dotted grey arrows).  (c), (d) and (e) are the best-fit parameters $r_1$, $C_2$, and $r_2$ determined by fitting simulated $g^{(2)}$(\(\tau\)) data using Equation 2 in the main text with \emph{N} = 2. (f) is the steady state PL saturation behaviour fitted by Equation 1 in the main text.  The results are plotted as a function of  $k_{12}$/$k_{21}$, where  $k_{21}$= \SI{200}{\mega Hz} is a fixed parameter.}
\label{Fig. S3}
\centering
\end{figure}

\newpage
\bibliography{references}

\end{document}